\documentclass[nature-comm,superscriptaddress,floatfix,tightenlines,showpacs,twocolumn]{revtex4-1}
\usepackage{graphicx}
\usepackage{dcolumn}   % needed for some tables
\usepackage{bm}        % for math
\usepackage{amssymb}   % for math
\usepackage{amsmath}
\usepackage{url}
\usepackage{layout}
\usepackage{color}
\usepackage{epsfig}
\usepackage{graphicx}
\usepackage{mathrsfs}\usepackage{bm}\usepackage{appendix}\usepackage{color}

\usepackage[thinlines]{easytable}
\usepackage{makecell}
\usepackage{tikz,pgf}
\usepackage{booktabs}
\usepackage{float}
\usepackage{hyperref}
\hypersetup{colorlinks,allcolors=blue}

\begin{document}

\title{Pairing without $\gamma$-Pocket in the La$_3$Ni$_2$O$_{7}$ Thin Film}

\author{Zhi-Yan Shao}
\thanks{These authors contributed equally to this work.}
\affiliation{School of Physics, Beijing Institute of Technology, Beijing 100081, China}

\author{Chen Lu}
\thanks{These authors contributed equally to this work.}
\affiliation{School of Physics, Hangzhou Normal University, Hangzhou 311121, China}

\author{Min Liu}
\thanks{These authors contributed equally to this work.}
\affiliation{College of Mathematics and Physics, Beijing University of Chemical Technology, Beijing 100029, China}

\author{Yu-Bo Liu}
\email{yuboliu@itp.ac.cn}
\affiliation{Institute of Theoretical Physics, Chinese Academic of Science, Beijing 100080, China}

\author{Zhiming Pan}
\affiliation{Department of Physics, Xiamen University, Xiamen 361005, Fujian, China}

\author{Congjun Wu}
\affiliation{New Cornerstone Science Laboratory, Department of Physics, School of Science, Westlake University, Hangzhou 310024, Zhejiang, China}
\affiliation{Institute for Theoretical Sciences, Westlake University, Hangzhou 310024, Zhejiang, China}
\affiliation{Key Laboratory for Quantum Materials of Zhejiang Province, School of Science, Westlake University, Hangzhou 310024, Zhejiang, China}
\affiliation{Institute of Natural Sciences, Westlake Institute for Advanced Study, Hangzhou 310024, Zhejiang, China}

\author{Fan Yang}
\email{yangfan\_blg@bit.edu.cn}
\affiliation{School of Physics, Beijing Institute of Technology, Beijing 100081, China}

\begin{abstract}
The recent discovery of high-temperature superconductivity (HTSC) in the La$_3$Ni$_2$O$_7$ ultrathin film at ambient pressure has aroused great research interest. The $\gamma$-pocket formed by the bonding $d_{z^2}$ band, which was previously proposed to be crucial in the pairing mechanism of pressurized bulk La$_3$Ni$_2$O$_7$, is reported to be either present or absent here by different experimental groups, giving rise to the problem: what is the pairing mechanism and pairing nature without the $\gamma$-pocket? Here, we start from a band structure obtained via density-functional-theoretical calculation, which exhibits no $\gamma$-pocket. Then, equipped with electron interactions, we study the pairing nature via combined weak- and strong- coupling approaches, which provide consistent results. In the weak-coupling study, the nesting between the $\alpha$- and $\beta$- pockets leads to an $s^\pm$-wave pairing in which the gap signs on the two pockets are opposite, as provided by our random-phase-approximation  based calculations. In real-space, the pairing pattern is dominated by the interlayer pairing of the $d_{x^2-y^2}$ orbital. In the strong-coupling study, as the $d_{z^2}$ orbitals are nearly half-filled and hence localized, the $d_{x^2-y^2}$ orbitals carry the HTSC. Driven by the interlayer superexchange transferred from the $d_{z^2}$ orbital through the Hund's rule coupling, the $d_{x^2-y^2}$ orbital electrons form interlayer $s$-wave pairing, as suggested by our slave-boson-mean-field study on the related two-orbital $t$-$J$ model. Projected onto the Fermi surface, this pairing just gives the $s^\pm$-wave pattern consistent with that obtained in the weak-coupling study. Our result is consistent with that obtained in recent scanning tunneling microscopy experiment.
\noindent

\end{abstract}\maketitle

~~~~~~~~~~~~~~~~~~~~~~~~~~~~~~~~~~~~~~~~~~~~~~~~~~~~~~~~~~~~~~~~~~~~~~~~~~~~~~~~~~~~~~~~~~~~~~~~~~~~~~~~~~~~~~~~~~~~~~~~~~~~~~~~~~~~~~~~~~~~~~~

\noindent 

% \section*{Introduction}
The discovery of high-temperature superconductivity (SC) (HTSC) in pressurized bulk La$_3$Ni$_2$O$_7$ \cite{Wang2023LNO, YuanHQ2023LNO,Wang2023LNOb,wang2023LNOpoly,wang2023la2prnio7,zhang2023pressure,zhou2023evidence,wang2024bulk,li2024pressure} has attracted wide interests, including both experimental \cite{YuanHQ2023LNO,Wang2023LNOb,wang2023LNOpoly,wang2023la2prnio7,zhang2023pressure,zhou2023evidence,wang2024bulk,li2024pressure,puphal2024unconven,Dong2024vis,Chen2024poly,wang2024chemical,Li2024ele,li2024distinguishing,zhou2024revealing,cui2023strain,yang2024orbital,wang2023structure,huo2025modulation,liu2024electronic} and theoretical studies \cite{cao2023flat,WangQH2023,YangF2023,lu2023bilayertJ,oh2023type2,qu2023bilayer,Yi_Feng2023,jiang2023high,qin2023high,zhang2023strong,pan2023rno,yang2023strong,fan2023sc,wu2024deconfined,zhang2024prediction,zhang2024s,Yang2024effective,zhang2024electronic,yang2024decom,Lu2024interplay,ZhangGM2023DMRG,heier2023competing,
sui2023rno,YaoDX2023,Dagotto2023,cao2023flat,zhang2023structural,huang2023impurity,geisler2023structural,rhodes2023structural,zhang2023la3ni2o6,yuan2023trilayer,li2024la3,geisler2024optical,li2017fermiology,wang2024non,chen2024tri,Li2024design,WangQH2023,YangF2023,lechermann2023,Kuroki2023,HuJP2023,lu2023bilayertJ,oh2023type2,liao2023electron,qu2023bilayer,Yi_Feng2023,jiang2023high,zhang2023trends,qin2023high,tian2023correlation,jiang2023pressure,lu2023sc,kitamine2023,luo2023high,zhang2023strong,pan2023rno,sakakibara2023La4Ni3O10,lange2023mixedtj,yang2023strong,lange2023feshbach,kaneko2023pair,fan2023sc,wu2024deconfined,zhang2024prediction,zhang2024s,Yang2024effective,zhang2024electronic,yang2024decom,ryee2024quenched,Lu2024interplay,Ouyang2024absence,ZhangGM2023DMRG,Werner2023,shilenko2023correlated,WuWei2023charge,chen2025charge,ouyang2023hund,heier2023competing,wang2024electronic,botzel2024theory,tian2024spin,Yubo_Liu2024,yin2025spmep,PhysRevB.111.104505,kaneko2025tj,Ji2025StrongCouplingLimit, Wang_2025,haque2025dft,shi2025theoretical,gao2025robust}. Previously, HTSC in nickelates was only found under high pressure (HP) but not found at ambient pressure (AP), which brings difficulties to studying the SC properties of nickelates since most of the experiments can only be conducted at AP. Recently, HTSC was found in the (La,Pr)$_3$Ni$_2$O$_7$ ultrathin film grown on the SrLaAlO$_4$ substrate at AP by two different groups independently \cite{Ko2024signature,zhou2024ambient,liu2025superconductivity}. This breakthrough attracts a lot of interests \cite{yue2025correlated,li2025photoemission,bhatt2025resolving,shao2025band,shi2025effect,le2025landscape,Daoxin_Yao2025,wang2025electronic,shao_electric_field,huang2025spm,geisler2025electronic,shen2025anomalous,hao2025superconductivity,fan2025superconducting,ushio2025theoretical,qiu2025pairing,sun2025observation,li2025enhanced,zhu2025quantum,cao2025strain}. 

There exists significant difference between the band structures of bulk La$_3$Ni$_2$O$_7$ at AP and under HP. The bonding Ni-$3d_{z^2}$ band crosses the Fermi level  forming the $\gamma$-pocket under HP while this band is below the Fermi level at AP \cite{Wang2023LNO}. Therefore, it was previously believed that the $\gamma$-pocket is crucial to SC \cite{WangQH2023,YangF2023,qin2023high,sui2023rno,zhang2023la3ni2o6,lechermann2023}. 
% 膜的有gamma的ARPES和理论，看起来没有gamma的ARPES，不知区别可能材料生长细节不同，都超导，看起来超导和gamma无关，没有gamma时是否超导如果有超导对称性和pairing nature
The $\gamma$-pocket is also observed in the (La,Pr)$_3$Ni$_2$O$_7$/SrLaAlO$_4$ ultrathin film at AP by angle-resolved photoemission spectroscopy (ARPES) experiment \cite{li2025photoemission}. Then some teams study the SC properties based on the band structure \cite{yue2025correlated,shao2025band,qiu2025pairing,cao2025strain}. 
However, some other ARPES results show that the top of the bonding Ni-$3d_{z^2}$ band is below the Fermi level in superconducting (La,Pr)$_3$Ni$_2$O$_7$/SrLaAlO$_4$ ultrathin film \cite{wang2025electronic,sun2025observation}, exhibiting no $\gamma$-pocket. These two conflicting experimental results seem to suggest that the SC in La$_3$Ni$_2$O$_7$ film is not intimately related to the presence of the $\gamma$-pocket. Since previous theoretical investigations on the pairing mechanism of this material are based on band structures hosting the $\gamma$-pocket, it is needed to explain here why SC can appear in the absence of the $\gamma$-pocket. 

In this paper, we first perform density functional theory (DFT) $+U$ calculation and obtain a band structure in which the Ni-$3d_{z^2}$ band is below the Fermi energy and then obtain a four-band tight-binding (TB) model. Based on this TB model, we study the SC properties both at the weak-coupling limit by random-phase-approximation (RPA) approach and at the strong-coupling limit by the slave-boson mean-field (SBMF) theory. 
% The results obtained from both methods are qualitatively consistent. 
% 主要结果。没有gamma也有超导
% Both approaches give the results that the system is superconducting and the dominant pairing is the interlayer $s_{\pm}$-wave pairing of the $d_{x^2-y^2}$ orbital. Since the bonding Ni-$3d_{z^2}$ band in the band structure considered here is below the Fermi level, our results show that the SC can appear even without the bonding Ni-$3d_{z^2}$ band crossing the Fermi level. 
The weak-coupling RPA results show that largest value in the distribution of the spin susceptibility is located at the nesting vector between the $\alpha$-pocket and the $\beta$-pocket, and the spin fluctuation of the momentum equals to the nesting vector mediates $s^{\pm}$-wave superconductivity, whose gap function on the $\alpha$-pocket and the one on the $\beta$-pocket have different signs. In the real-space pairing configuration, the interlayer pairing of the $d_{x^2-y^2}$ electrons is dominant. In the strong-coupling limit, the absence of the $\gamma$-pocket dictates that the $d_{z^2}$ orbital is nearly exactly half-filled and hence localized, leaving the $d_{x^2-y^2}$ orbital as the charge carrier.  The strong Hund's coupling effectively transmits the interlayer superexchange interaction $J_{\perp}$ associated with the localized $d_{z^2}$ orbital to the $d_{x^2 - y^2}$ orbital, thereby inducing an interlayer pairing channel in the latter. Our SBMF theory suggests that, the interlayer pairing of the $d_{x^2 - y^2}$ orbital emerges as the dominant superconducting channel. By projecting the superconducting gap onto the Fermi surface, we find that the gap function exhibits a sign change between the $\alpha$- and $\beta$-pockets, indicating an $s^{\pm}$-wave pairing symmetry. This is qualitatively consistent with the results obtained in the weak-coupling limit. Our pairing gap function is consistent with that observed in the scanning tunneling microscopy (STM) experiment.

\section*{Results}
\begin{figure}[htbp]
    \centering
    \includegraphics[width=1\linewidth]{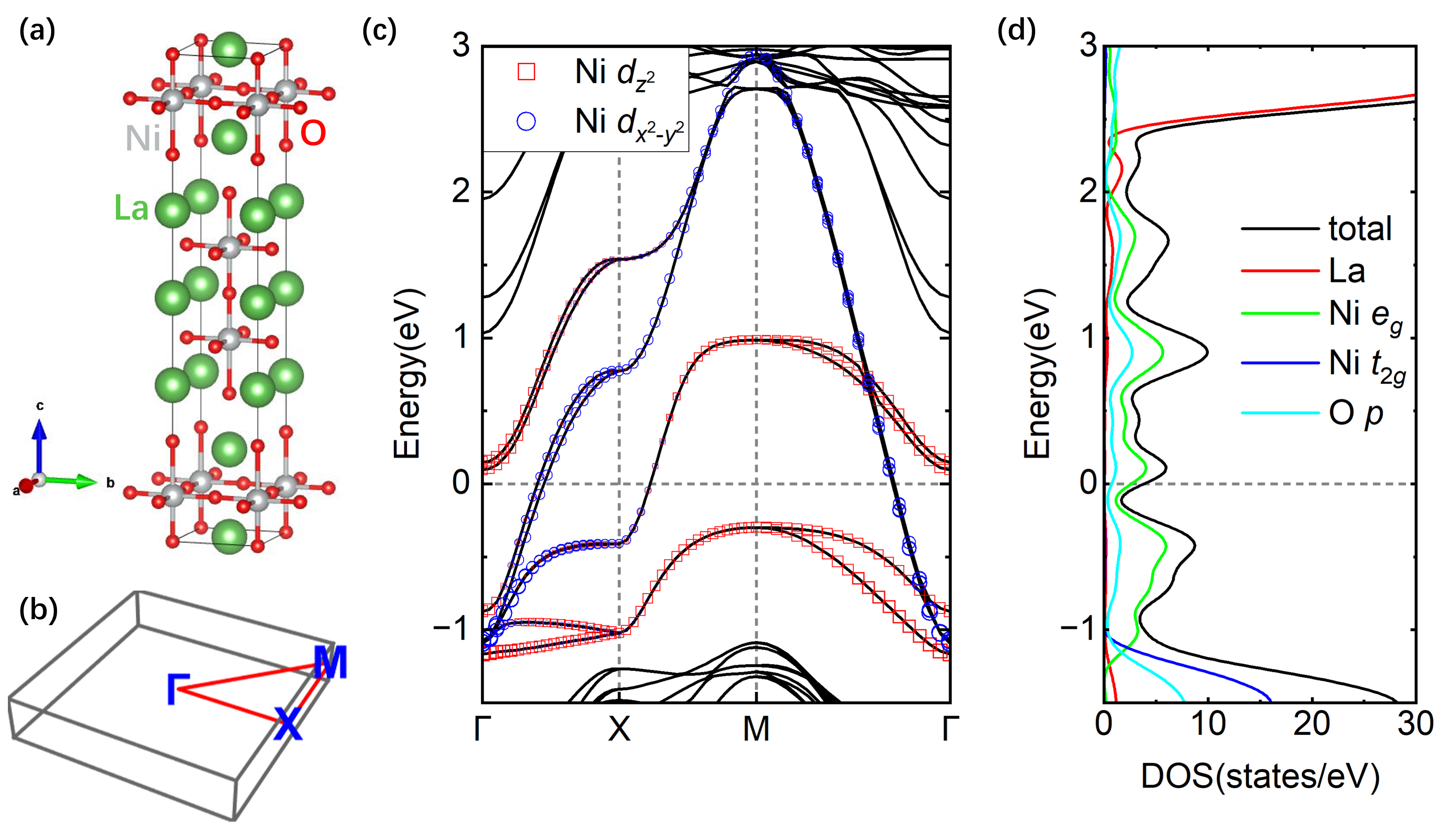}
    \caption{(a) The optimized crystal structure. (b) BZ with high-symmetric points and lines marked. (c) The DFT$+U$ band structure with $U=3.5~\mathrm{eV}$. The red and blue markers represent the weight of Ni-$d_{z^2}$ and Ni-$d_{x^2-y^2}$ orbitals, respectively. (d) The total DOS and the DOS of La, Ni-$d$ orbitals and O-$p$ orbitals. } \label{fig_structure_bz_band_dos}
\end{figure}

% \subsection{The Density Functional Theory Study and the Tight-Binding Model}
The band structure of the La\textsubscript{3}Ni\textsubscript{2}O\textsubscript{7} film grown on the SrLaAlO\textsubscript{4} substrate was performed using DFT$+U$ method. The computational details are provided in \textbf{METHODS}. 
The optimized crystal structure (Fig. \ref{fig_structure_bz_band_dos} (a)) has the $I4/mmm$ symmetry, where the lattice constants are fixed as observed in experiment, i.e. $a=3.756~\mathrm{\AA}$ and $c=20.72~\mathrm{\AA}$ \cite{Ko2024signature,wang2025electronic}, and the atomic positions are fully relaxed. 
In this optimized crystal structure, the Ni-O-Ni angle between the NiO layers is $180^{\circ}$. For a single-bilayer, the unit cell includes two Ni atoms with one in the top layer and the other in the bottom layer. 
The band structure along the high-symmetric line in the Brillouin zone (BZ) (Fig. \ref{fig_structure_bz_band_dos} (b)) and the density of state (DOS) are shown in Fig. \ref{fig_structure_bz_band_dos} (c) and (d), respectively. The band and DOS show that the Ni-$3d$-$e_g$ orbitals are mainly located below the Fermi level. Particularly, there is an energy gap about 300 meV of the bonding $d_{z^2}$ band below the Fermi energy at M point. In addition, the $d_{z^2}$ band does not cross the Fermi level at the $\Gamma$ point, and thus forms no electron pocket. This feature is different from the band structure of the pressurized bulk La\textsubscript{3}Ni\textsubscript{2}O\textsubscript{7} that the bonding $d_{z^2}$ band crosses the Fermi energy. 

\begin{figure}[htbp]
    \centering
    \includegraphics[width=1\linewidth]{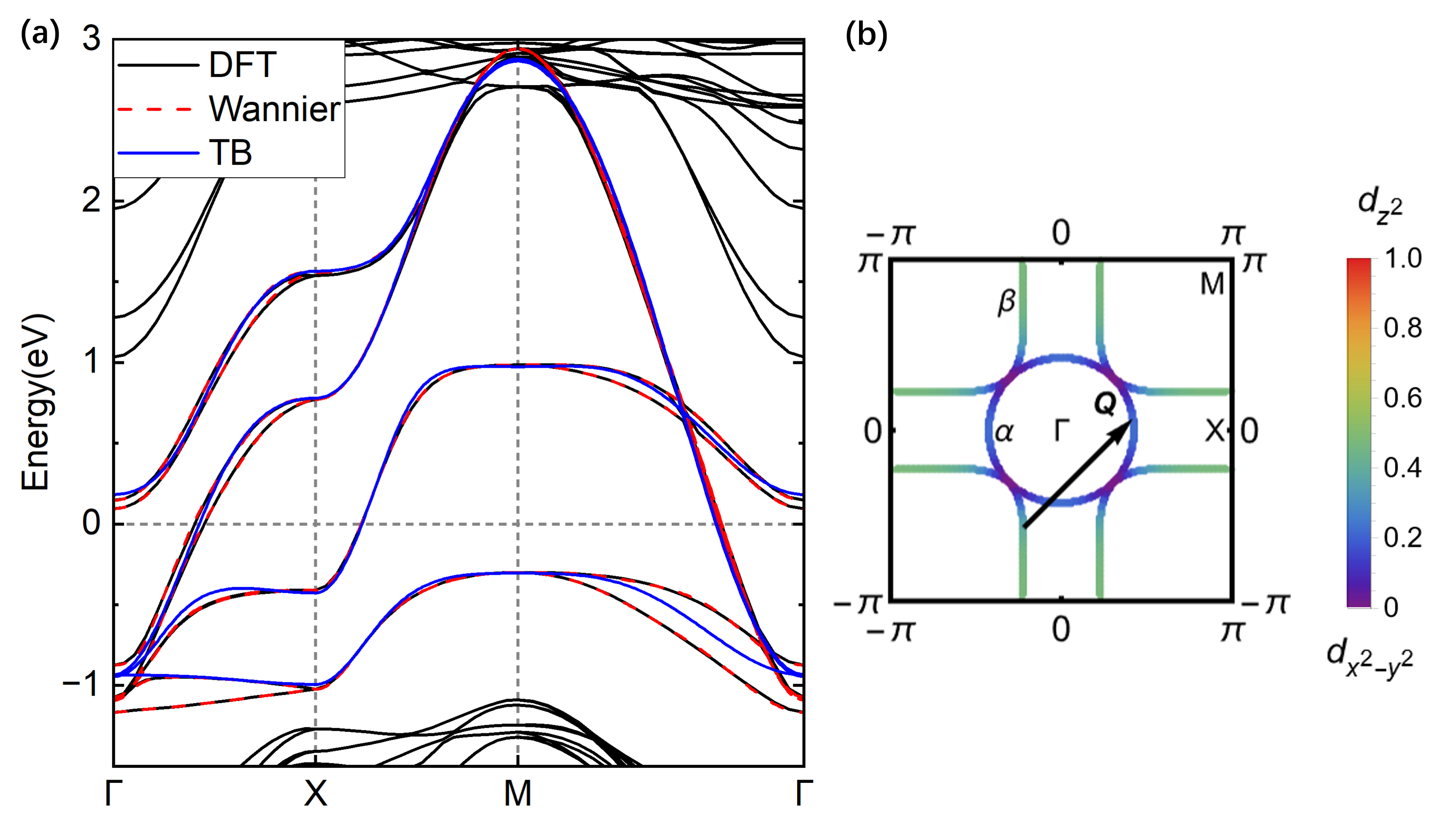}
    \caption{(a) The DFT band (black line), the Wannier band (red dashed line) and the TB band (blue line). They almost coincide with each other near the Fermi energy (horizontal gray dashed line). (b) The FS obtained from the TB Hamiltonian. The pockets are labeled as $\alpha$ and $\beta$. $\bm{Q}$ marks the nesting vector. } \label{fig_band_fs}
\end{figure}

For convenience in the subsequent studies, we construct a Wannier model with the Ni-$d_{z^2}$ and Ni-$d_{x^2-y^2}$ orbitals based on the maximally-localized Wannier function method implemented in the WANNIER90 code \cite{mostofi2008wannier90}. Then, from the Wannier model, we obtain a TB Hamiltonian describing the hopping between the Ni-$d_{z^2}$ and Ni-$d_{x^2-y^2}$ orbitals within a single-bilayer, whose expression is given in \textbf{METHODS}. 
This TB model has $D_4$ symmetry. 
Each unit cell includes two Ni atoms, one in the top layer and the other in the bottom layer. 
The Wannier bands and the TB bands almost coincide with the DFT band near the Fermi energy, as shown in Fig. \ref{fig_band_fs} (a).  
In this TB model, the intralayer nearest neighbor hopping of the $d_{x^2-y^2}$ electrons $t_{1}^{x}=-0.4765~\mathrm{eV}$ is similar to the ones in the previous TB models \cite{yue2025correlated,shao2025band}, but the interlayer hopping of the $d_{z^2}$ electrons $t_{\perp}^{z}=-0.6408~\mathrm{eV}$ is larger than the ones in the previous TB models \cite{yue2025correlated,shao2025band}. This larger $t_{\perp}^{z}$ leads to the larger splitting between the bonding Ni-$3d_{z^2}$ band and the anti-bonding Ni-$3d_{z^2}$ band and the gap between the top of the bonding Ni-$3d_{z^2}$ band and the Fermi level. 
Fig. \ref{fig_band_fs} (b) shows the Fermi surface (FS) obtained from the TB model. An electron pocket $\alpha$ centering around the $\Gamma$ point and a hole pocket $\beta$ centering around the M point exist, which is consistent with the ARPES results reported in Ref. \cite{wang2025electronic,sun2025observation}. Both $\alpha$- and $\beta$- pockets show mixing of orbital contents. The dominant nesting is between the $\alpha$-pocket and the $\beta$-pocket with nesting vector $\bm{Q}\approx(0.64\pi,0.64\pi)$. 

\subsection{Weak-Coupling Study}

After considering the interaction between electrons, we obtain the multi-orbital Hubbard Hamiltonian, 
\begin{equation} \label{eq_Hubbard}
    \begin{aligned}
       H_{\mathrm{H}}=& H_{\mathrm{TB}}+U \sum_{i\mu\alpha} n_{i\mu\alpha\uparrow}n_{i\mu\alpha\downarrow} + V \sum_{i\mu} n_{i\mu{}z}n_{i\mu{}x} \\ & + J_H \sum_{i\mu} \left[ \left( \sum_{\sigma\sigma^{\prime}} c^{\dagger}_{i\mu{}z\sigma}c^{\dagger}_{i\mu{}x\sigma^{\prime}}c_{i\mu{}z\sigma^{\prime}}c_{i\mu{}x\sigma} \right) \right. \\ & \phantom{+J_H\sum_{i\mu}} \left. + \left( c^{\dagger}_{i\mu{}z\uparrow}c^{\dagger}_{i\mu{}z\downarrow}c_{i\mu{}x\downarrow}c_{i\mu{}x\uparrow} + \mathrm{h.c.} \right) \right], 
    \end{aligned}
\end{equation}
where the $U$ and $V$ terms describe the intraorbital and interorbital repulsions, respectively, and the $J_H$ term describes the Hund's coupling and pair hopping. We use the relation $U=V+2J_H$ and set $J_H=U/6$ here. We study this model at weak-coupling limit through the standard multi-orbital RPA approach \cite{takimoto2004strong,yada2005origin,kubo2007pairing,graser2009near,liu2013d+,zhang2022lifshitz,kuroki101unconventional}, which is briefly introduced in \textbf{METHODS}. 

\begin{figure}[htbp]
    \centering
    \includegraphics[width=1\linewidth]{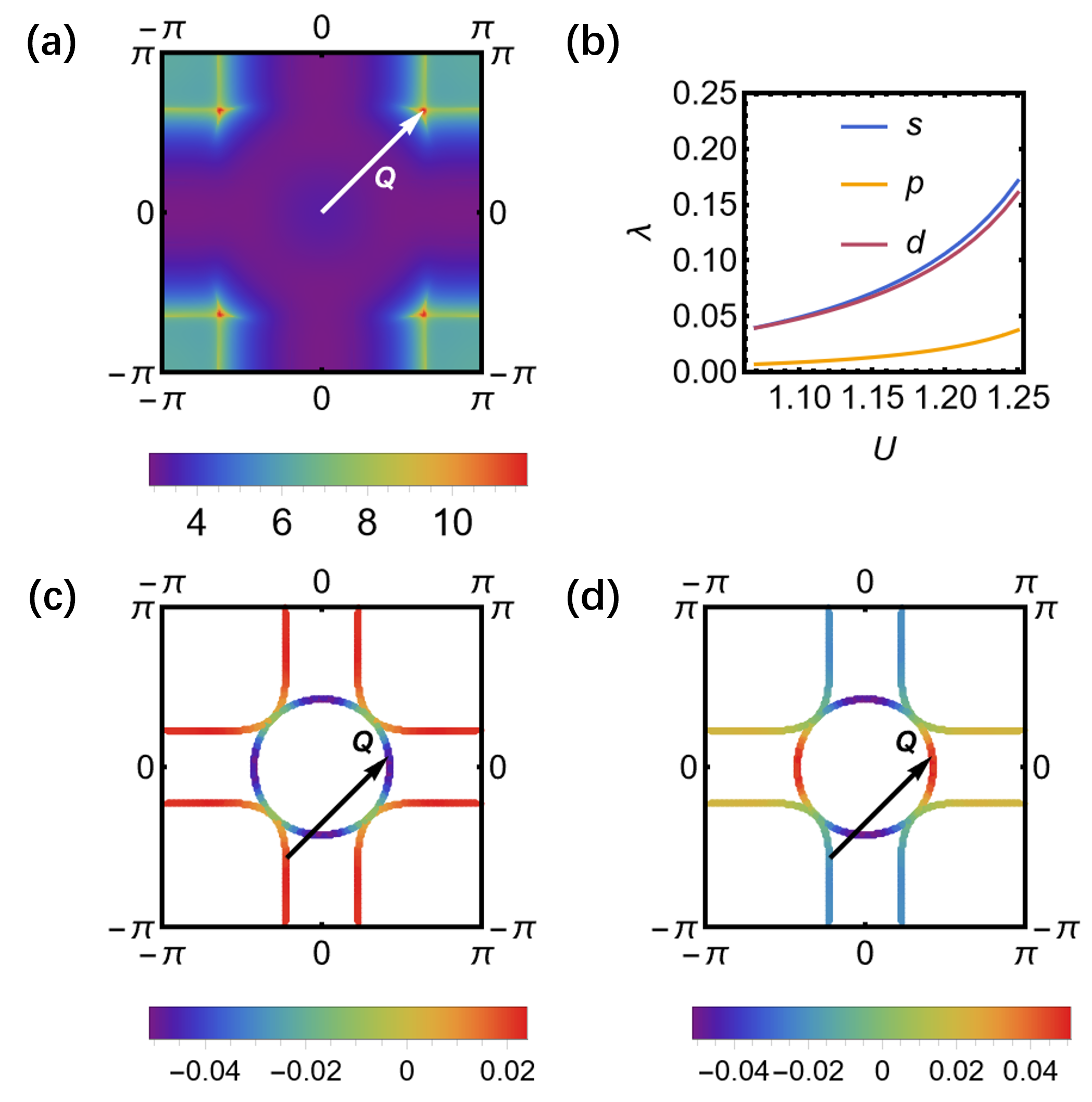}
    \caption{(a) The distribution of the spin susceptibility over the BZ for $U=1.25~\mathrm{eV}<U_c$. The largest value is located at $\bm{Q}\approx(0.64\pi,0.64\pi)$, which is just the nesting vector of the FS. (b) The pairing eigenvalue $\lambda$ as function of interaction strength $U$ for different pairing symmetries (marked by different colors). (c)-(d) The distribution of the leading $s^{\pm}$-wave pairing gap function and the subleading $d_{x^2-y^2}$-wave pairing gap function on the FS for $U=1.25~\mathrm{eV}$, respectively. } \label{fig_rpa_chi_lambda_gap}
\end{figure}

The RPA results are shown in Fig. \ref{fig_rpa_chi_lambda_gap}. Fig. \ref{fig_rpa_chi_lambda_gap} (a) shows the distribution of the spin susceptibility $\chi^{(s)}(\bm{k})$ for $U=1.25~\mathrm{eV}<U_c=1.33~\mathrm{eV}$. The largest value is located at $\bm{Q}\approx(0.64\pi,0.64\pi)$, which is just the nesting vector of the FS. 

The pairing eigenvalue $\lambda$ as function of interaction strength $U$ for different pairing symmetries is shown in Fig. \ref{fig_rpa_chi_lambda_gap} (b). $\lambda$ for all pairing symmetries increases as $U$ increases. For considered $U$ values, the $s$-wave pairing dominates over other pairing symmetries, and $\lambda$ for the $d$-wave pairing is slightly smaller than the one for the $s$-wave pairing. The distribution of the $s$-wave pairing and the $d$-wave pairing is shown in Fig. \ref{fig_rpa_chi_lambda_gap} (c) and (d), respectively. In Fig. \ref{fig_rpa_chi_lambda_gap} (c), the gap function has the $s^{\pm}$ symmetry that it remains the same under any symmetric transformations and the signs of the gap function on the $\alpha$- and $\beta$- pockets are opposite. In this $s^{\pm}$-wave gap function, the value near the momentum that the $\alpha$- and $\beta$- pockets touch each other is relatively small but not zero, i.e., no gap node is located near the momentum. In Fig. \ref{fig_rpa_chi_lambda_gap} (d), the gap function has the $d_{x^2-y^2}$ symmetry that it changes sign with every $90^{\circ}$ rotation or upon the mirror reflection about the diagonals of the BZ. The value of the gap function is zero along the diagonals of the BZ due to the restriction of the $d_{x^2-y^2}$ symmetry. This is inconsistent with the ARPES results reported in Ref. \cite{shen2025anomalous} that no gap node is detected near the momentum that the $\alpha$- and $\beta$- pockets touch each other. Thus, the pairing symmetry of the system should be $s^{\pm}$. 

\begin{figure}[htbp]
    \centering
    \includegraphics[width=1\linewidth]{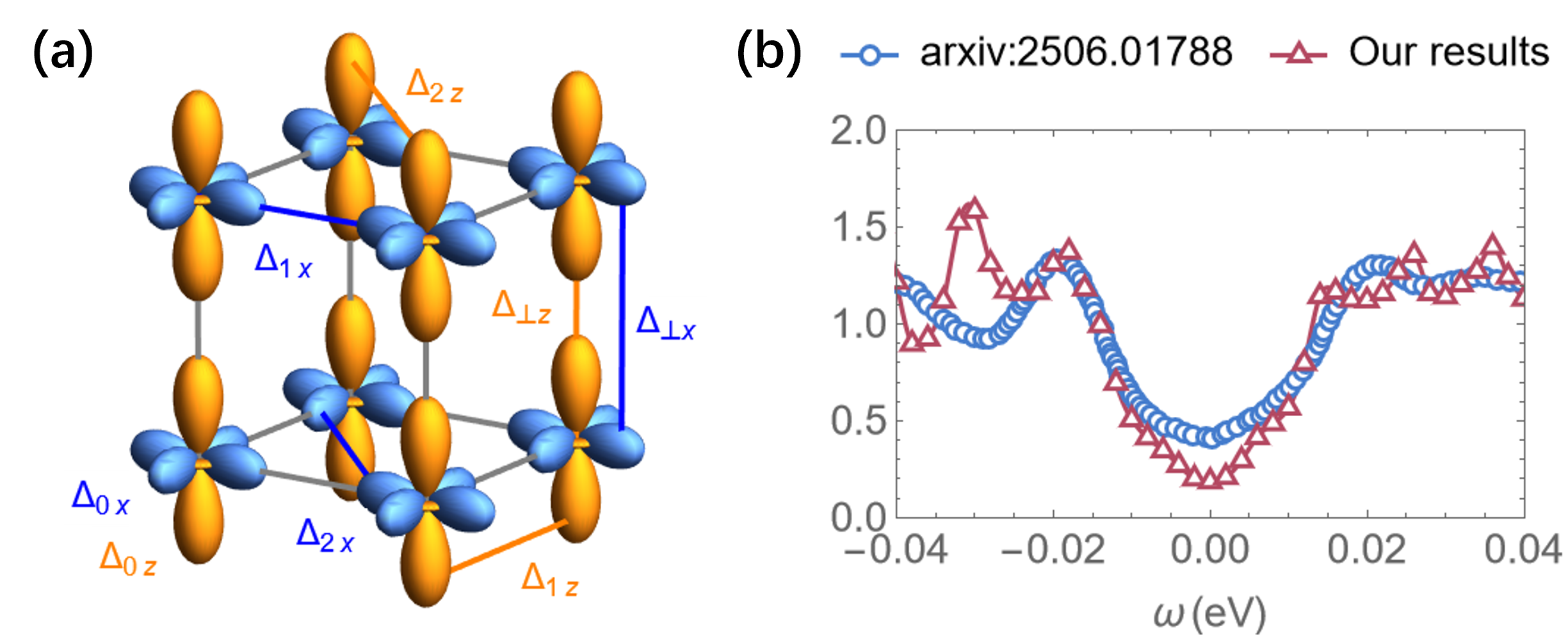}
    \caption{(a) Illustration of the real-space pairing configuration. (b) The theoretical STM spectrum (red), compared with the experimental results in Ref. \cite{fan2025superconducting} (blue). } \label{fig_rpa_gap_real_stm}
\end{figure}

We further study the nature of the $s^{\pm}$-wave pairing. The real-space pairing configuration is illustrated in Fig. \ref{fig_rpa_gap_real_stm} (a). 
% The dominant pairing amplitude is the one of the interlayer pairing of the $d_{x^2-y^2}$ orbital, $\Delta_{\perp{}x}$. 
Some pairing amplitudes, in units of the one of the interlayer pairing of the $d_{x^2-y^2}$ orbital $\Delta_{\perp{}x}$, are listed as follows. 
The pairing amplitude of the interlayer pairing of the $d_{z^2}$ orbital $\Delta_{\perp{}z}=0.42\Delta_{\perp{}x}$. 
% The onsite intraorbital pairing amplitudes and the pairing amplitudes on other bonds are relatively small and are not illustrated. % 多列举一些
The onsite intraorbital pairing amplitudes of the $d_{x^2-y^2}$ orbital and the $d_{z^2}$ orbital are $\Delta_{0x}=0.047\Delta_{\perp{}x}$ and $\Delta_{0z}=-0.22\Delta_{\perp{}x}$, respectively. The nearest neighbor intralayer pairing amplitudes of the $d_{x^2-y^2}$ orbital and the $d_{z^2}$ orbital are $\Delta_{1x}=0.24\Delta_{\perp{}x}$ and $\Delta_{1z}=0.036\Delta_{\perp{}x}$, respectively. The next nearest neighbor intralayer pairing amplitudes of the $d_{x^2-y^2}$ orbital and the $d_{z^2}$ orbital are $\Delta_{2x}=0.3\Delta_{\perp{}x}$ and $\Delta_{2z}=0.17\Delta_{\perp{}x}$, respectively. The pairing amplitudes of the other bonds are relatively small and are not illustrated. Above all, the dominant pairing channel is the interlayer pairing of the $d_{x^2-y^2}$ orbital, $\Delta_{\perp{}x}$. 

We also calculate the STM spectrum of the $s^{\pm}$-wave pairing state. 
We consider the Matsubara Green's function
\begin{equation}
    \mathcal{G}(i,\tau) = - \sum_{\mu\alpha\sigma} \left\langle T_{\tau} c_{i\mu\alpha\sigma}(\tau) c^{\dagger}_{i\mu\alpha\sigma}(0) \right\rangle, 
\end{equation}
where $i$ is the site index, $\tau$ is the imaginary time, and $T_{\tau}$ represents the time-ordered product of the operators. 
Then the spectrum can be obtained from the imaginary part of the analytic continuation of $\mathcal{G}(i,\mathrm{i}\omega_n)$, $\mathcal{G}(i,\mathrm{i}\omega_n\rightarrow\omega+\mathrm{i}0^+)$, where $\mathcal{G}(i,\mathrm{i}\omega_n)$ is the Fourier transformation of $\mathcal{G}(i,\tau)$. 
% The spectrum is obtained from the imaginary part of the retarded Green's function of the SC state, 
% \begin{equation}
%     - \frac{1}{\pi} \mathrm{Im}\left[ G^r(i,\omega+\mathrm{i}0^+) \right]. 
% \end{equation}
% Here, $i$ is the site index, $G^r(i,\omega+\mathrm{i}0^+)$ is the Fourier transform of $G^r(i,t)$, 
% \begin{equation}
%     G^r(i,\omega+\mathrm{i}0^+) = \int_{-\infty}^{\infty} \mathrm{d}t G^r(i,t) \mathrm{e}^{\mathrm{i}(\omega+\mathrm{i}0^+)t}, 
% \end{equation}
% and $G^r(i,t)$ is defined as 
% \begin{equation}
%     G^r(i,t) = -\mathrm{i} \theta(t) \sum_{\mu\alpha\sigma} \left\langle \left\{ c_{i\mu\alpha\sigma}(t), c^{\dag}_{i\mu\alpha\sigma}(0) \right\} \right\rangle. 
% \end{equation}
% Here, $\{\cdots\}$ is the anticommutator of two operators, and $\theta(t)$ is the unit step function. 
% \begin{equation}
%     -\mathrm{Im} \left[\sum_{\bm{k}m}\frac{u_{\bm{k}m}^2}{\omega-E_{\bm{k}m}+\mathrm{i}0^+}+\frac{v_{\bm{k}m}^2}{\omega+E_{\bm{k}m}+\mathrm{i}0^+}\right]
% \end{equation}
% where $m$ is the band index, $E_{\bm{k}m}=\sqrt{\left(\varepsilon_{\bm{k}m}-\mu\right)^2+\Delta_{\bm{k}m}^2}$, $\varepsilon_{\bm{k}\alpha}$ is the $m$-th eigenvalue of the $\bm{k}$-space TB Hamiltonian $H_{\bm{k}}$, $\mu$ is the chemical potential, $\Delta_{\bm{k}m}$ is the gap function of the momentum $\bm{k}$ and band $m$, $u_{\bm{k}m}^2=\frac{1}{2}\left(1+\frac{\varepsilon_{\bm{k}m}-\mu}{E_{\bm{k}m}}\right)$ and $v_{\bm{k}m}^2=\frac{1}{2}\left(1-\frac{\varepsilon_{\bm{k}m}-\mu}{E_{\bm{k}m}}\right)$. 
The results, compared with the experimental results in Ref. \cite{fan2025superconducting}, are shown in Fig. \ref{fig_rpa_gap_real_stm} (b). The shape of the theoretical curve is similar to the experimental one, indicating the possibility of the $s^{\pm}$-wave pairing. % 增加一些内容

\subsection{Strong-Coupling Study}
Due to the strong-correlated characteristic of the low-energy Ni-$3d$ orbitals, a strong-coupling point of view is also a reasonable starting point. In the strong-coupling limit, the onsite Coulomb interaction $U, V$ and the Hund’s coupling $J_H$ is no longer explicitly incorporated in the Hamiltonian. Instead, the role of these strong interactions is to bring constraint to the low-energy Hilbert space and to lead to low-energy effective superexchange interactions.

\begin{figure}[htbp]
\centering
\includegraphics[width=0.98\linewidth]{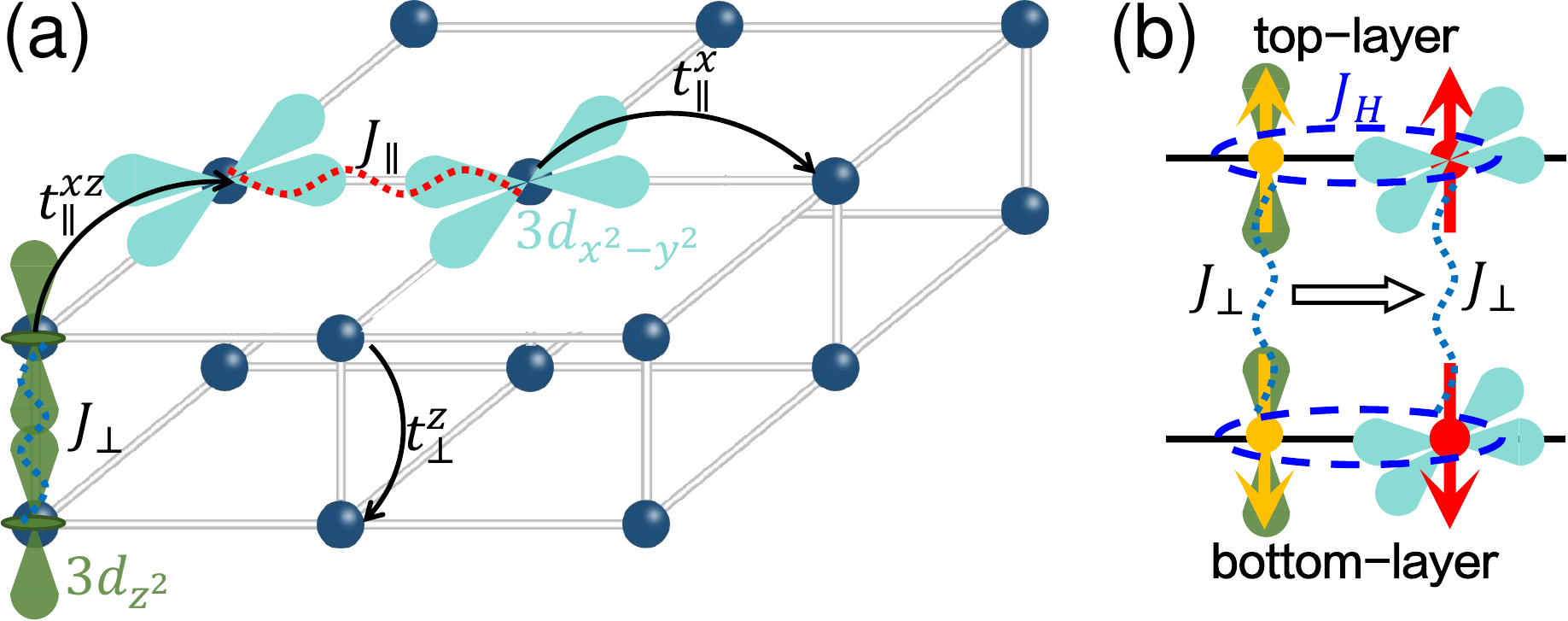}
    \caption{(a) Schematic illustration of the interlayer superexchange transmission between two orbitals mediated by Hund’s coupling. (b) Schematic illustration of the effective two-orbital $t$-$J$ model.} \label{model}
\end{figure}

As illustrated in Fig.~\ref{model}(a), the low-energy degrees of freedom are mainly contributed by the two $E_g$-orbitals. In the large-$U$ limit, each $E_g$ orbital can only be occupied by one electron or a hole, satisfying the ``no-double-occupancy'' constraint. The $3d_{x^2-y^2}$ orbital wave function is predominantly extended within the plane and thus this orbital features a sizable in-plane hopping amplitude $t^x_{\parallel}$, giving rise to an in-plane superexchange interaction $J_{\parallel}$. In contrast, the $3d_{z^2}$ orbital extends primarily along the out-of-plane direction, exhibiting a large interlayer hopping amplitude $t^z_{\perp}$, which leads to an interlayer superexchange interaction $J_{\perp}$. In addition, there exists a significant nearest-neighbor in-plane hybridization term $t^{xz}_{\parallel}$ between the two orbitals within the plane.

Due to the strong Hund’s coupling, electrons occupying different $E_g$ orbitals at the same site tend to align their spins. As a consequence, the interlayer $J_{\perp}$ generated in the $3d_{z^2}$ sector is transmitted to the $3d_{x^2 - y^2}$ channel, mediating an effective interlayer antiferromagnetic exchange between $3d_{x^2 - y^2}$ electrons~\cite{lu2023bilayertJ,Lu2024interplay}, as schematically illustrated in Fig.~\ref{model}(b). These considerations motivate the construction of the following effective two-orbital bilayer $t$–$J$ model~\cite{Lu2024interplay}
\begin{equation}
\begin{aligned} 
&H=H_{\mathrm{TB}}+H_{J,x}+H_{J,z},\\
&H_{J,x}=J_{\parallel} \sum_{\langle i,j\rangle \mu} \bm{S}_{\mu ,x}(i) \cdot \bm{S}_{\mu ,x}(j) \\
& \qquad+J_{\perp} \sum_{i} \bm{S}_{t,x}(i) \cdot \bm{S}_{b,x}(i),\\
&H_{J,z}=J_{\perp} \sum_{i} \bm{S}_{t,z}(i) \cdot \bm{S}_{b,z}(i).
\label{eq:xz-t-H1}
\end{aligned}
\end{equation}
where $\mu = t(\text{top}), b(\text{bottom})$ denotes the layer index, and $\bm{S}_{\mu ,\alpha}(i)$ is the spin operator for orbital $\alpha = x (d_{x^2-y^2})$ or $z(d_{z^2})$ at site $i$ on layer $\mu$. Due to the relation $J\propto  t^2$ in the strong-coupling limit, we adopt $J_{\perp}\approx 2J_{\parallel}\approx 0.4$eV.

\begin{figure}[htbp]
\centering
\includegraphics[width=1\linewidth]{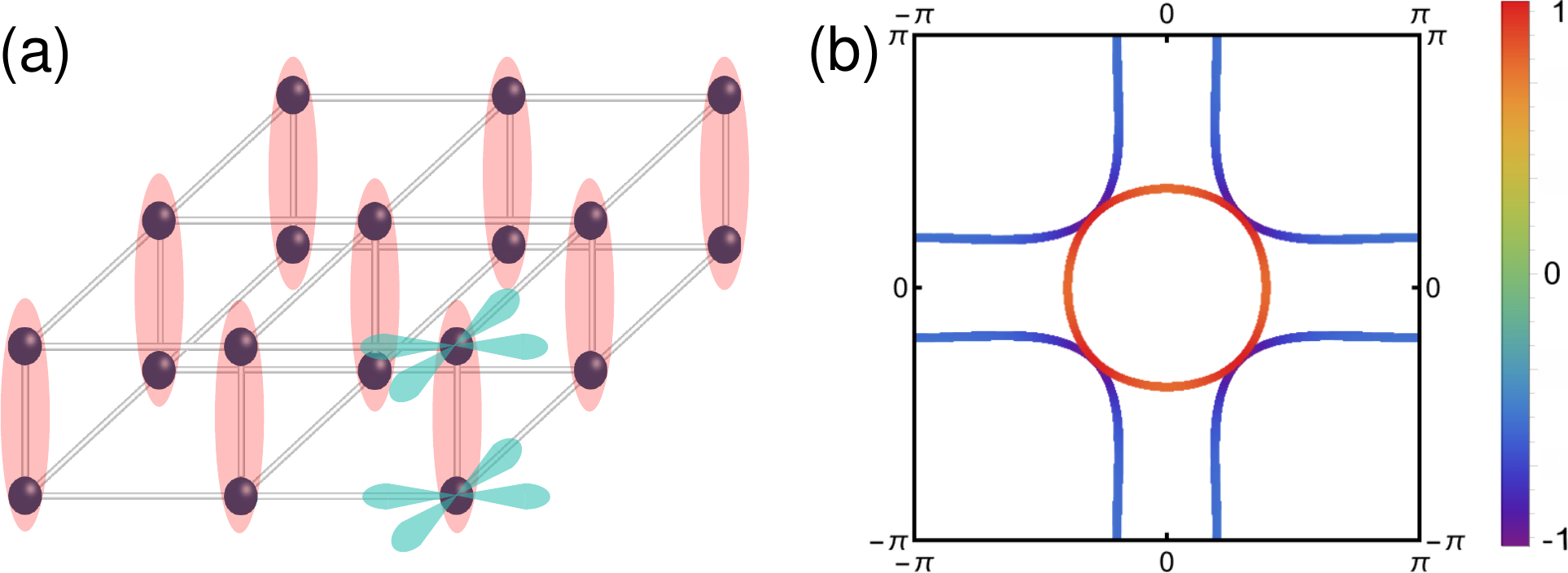}
    \caption{(a) Pairing configuration of the obtained interlayer-$s$-wave pairing for $d_{x^2 - y^2}$ orbital. (b) Projection of the superconducting gap $\delta_x \Delta^{x}_{\perp}$ onto the Fermi surface.} \label{SB_gap}
\end{figure}

We apply the SBMF approach on this effective $t$-$J$ model to investigate the SC properties for the film system. In the slave boson approach ~\cite{kotliar1988,lee2006htsc}, 
the electron operators of two $E_g$ orbitals are represented as $c_{\mu\alpha\sigma}^{\dagger}(i)=f_{\mu\alpha\sigma}^{\dagger} (i) b_{\mu\alpha}(i)$, where $f_{\mu\alpha\sigma}^{\dagger} (i)$ is the spinon creation operator and $b_{\mu\alpha}(i)$ is the holon annihilation operator for the two orbitals ($\alpha=x,z$) respectively. We focus on the ground-state in which the holons are in the Bose-Einstein-condensation, under which the holon operator is replaced by its expectation value $\sqrt{\delta_{\alpha}}$, where $\delta_{\alpha}$ denotes the hole doping level of the $\alpha$-orbital. The computational details are provided in {\bf{METHODS}}, and the complete set of obtained mean-field spinon order parameters is given in the Supplementary Material. 

Our calculations reveal that the three dominant spinon pairing amplitudes are as follows: the interlayer pairing of the $d_{z^2}$ spinons, $\Delta^{z}_{\perp} = 1.51 \times 10^{-1}$eV; the interlayer pairing of the $d_{x^2 - y^2}$ spinons, $\Delta^{x}_{\perp} = 1.28 \times 10^{-3}$eV; and the intralayer pairing of the $d_{x^2 - y^2}$ spinons, $\Delta^{x}_{\parallel} = 1.11 \times 10^{-5}$eV. Note that these spionon pairing amplitudes are unifiedly defined as $\Delta^{\alpha}_{\text{spinon}}=\frac{3}{8} J\langle f_{\alpha\downarrow}(j) f_{\alpha\uparrow} (i)-f_{\alpha\uparrow}(j) f_{\alpha\downarrow}(i) \rangle$, where $\alpha$ labels pairing channels. Obviously, the interlayer pairing dominates the intralayer one. Moreover, the real-space structure of the pairing respects all point group symmetries of the system, indicating an $s$-wave pairing. Within the SBMF framework, the superconducting order parameter in the channel $\alpha$ is given by $\Delta^{\alpha}_{\mathrm{SC}}\sim \left \langle c_{\alpha\uparrow }c_{\alpha\downarrow }  \right \rangle \sim  \delta_\alpha \Delta^{\alpha}_{\mathrm{spinon}}$. Due to absence of the $\gamma$-pocket formed mainly by the $d_{z^2}$ orbital, we have $\delta_z=0$, resulting in $\Delta^{z}_{\mathrm{SC}}=0$. This behavior can be attributed to the half-filled nature of the $d_{z^2}$ orbital, which suppresses charge fluctuations and thereby prevents the establishment of phase coherence in that channel. Consequently, the dominant superconducting channel becomes the interlayer pairing of the $d_{x^2 - y^2}$ orbitals, as illustrated in Fig.~\ref{SB_gap} (a).

Furthermore, to elucidate the $\bm{k}$-space nature of the superconducting pairing, we project the dominant superconducting gap $\delta_x \Delta^{x}_{\perp}$ onto the Fermi surface, as shown in Fig.~\ref{SB_gap} (b). The resulting gap structure is consistent with the weak-coupling RPA results, exhibiting a clear sign change between the $\alpha$- and $\beta$- pockets—characteristic of an $s^\pm$-wave pairing symmetry. Therefore, both our weak- and strong-coupling calculations consistently yield an $s^\pm$-wave superconducting state.

To compare the superconducting properties between the film at AP and the bulk under HP, we further carried out calculations for the bulk system under 30 GPa (numerical results provided in the Supplementary Material). In both cases, we find an $s$-wave superconducting state dominated by interlayer pairing in the $d_{x^2 - y^2}$ orbital. Notably, the superconducting gap in the film system at AP is significantly smaller than that in the bulk under HP, consistent with experimental observations. This difference can be attributed to two main factors. Firstly, the overall hopping amplitudes in the film at AP are smaller compared to those in the bulk system at 30 GPa, leading to reduced superexchange interactions and consequently a smaller superconducting gap. Secondly, the film lacks the $\gamma$-pocket that is present in the bulk, reducing the hole-doping level of the $d_{z^2}$ orbital. Consequently, the filling fraction of the $d_{x^2 - y^2}$ orbital is reduced, and this orbital becomes more overdoped, which in turn weakens its pairing strength.

\section*{Discussion}
We study the SC properties of the La$_3$Ni$_2$O$_7$ ultrathin film grown on the SrLaAlO$_4$ substrate. We obtain a band structure without the $\gamma$-pocket by DFT$+U$ calculations. Based on this band structure, we consider electron-electron interaction-driven pairing mechanism and pairing nature, both at the weak-coupling limit by the standard multi-orbital RPA approach and at the strong-coupling limit by the SBMF theory. Both approaches give the results of the $s^{\pm}$-wave pairing without gap node, mainly contributed by the interlayer pairing of the Ni-$d_{x^2-y^2}$ electrons. Furthermore, the calculated STM spectrum is similar to the experimental results reported in Ref. \cite{fan2025superconducting}. Our results suggest that the presence of the $\gamma$-pocket is not necessary to the pairing mechanism of the bilayer nickelates. 

% no gamma pocket -> lower Tc

\section*{Methods}

\subsection{Density Functional Theory Calculations}

Vienna ab inito simulation package (VASP) is used in our DFT$+U$ calculations, where the projector augmented wave (PAW) is employed as the pseudopotentials \cite{kress1993vasp,kress1996planewave,blochl1994paw}. The electronic correlations are considered by the generalized gradient approximation (GGA) and the Perdew-Burke-Ernzerhof (PBE) exchange potential \cite{perdew1996gga}. A $k$-point grid $8 \times 8 \times 2$ is adopted. The plane-wave cutoff energy is set as $520~\mathrm{eV}$. We set a reasonable onsite Coulomb interaction $U=3.5~\mathrm{eV}$ \cite{Wang2023LNO,yang2024orbital} in our DFT$+U$ calculations. 

\subsection{Tight-Binding Model}

The TB Hamiltonian describing the hopping within a single-bilayer can be expressed as 
\begin{equation} \label{eq_TB_general_real}
    H_{\mathrm{TB}} = \sum_{i,j,\mu,\nu,\alpha,\beta,\sigma} t_{i\mu\alpha,j\nu\beta} c^{\dagger}_{i\mu\alpha\sigma} c_{j\nu\beta\sigma}. 
\end{equation}
Here, $i$ and $j$ label sites, $\mu,\nu=t,b$ represent the top ($t$) or bottom ($b$) layer, $\alpha,\beta=z,x$ represent the Ni-$d_{z^2}$ ($z$) or Ni-$d_{x^2-y^2}$ ($x$) orbital, and $\sigma$ is the spin of an electron. The TB Hamiltonian can be transformed into the $\bm{k}$-space, 
\begin{equation} \label{eq_TB_general_k}
    H_{\mathrm{TB}} = \sum_{\bm{k},\mu,\nu,\alpha,\beta,\sigma} \left(H_{\bm{k}}\right)_{\mu\alpha,\nu\beta} c^{\dagger}_{\bm{k}\mu\alpha\sigma} c_{\bm{k}\nu\beta\sigma}. 
\end{equation}
Here, $H_{\bm{k}}$ is the $\bm{k}$-space Hamiltonian matrix. In the basis $(tz, tx, bz, bx)$, it can be expressed as 
\begin{equation}
    H_{\bm{k}} = \left(
    \begin{array}{cc}
        H_1 & H_2 \\
        H_2^{\dagger} & H_1
    \end{array}
    \right).
\end{equation}
Here, the blocks $H_1$ and $H_2$ are expressed as 
\begin{equation}
    H_1 = \left(
    \begin{array}{cc}
        h_1^z & h_2 \\
        h_2 & h_1^x
    \end{array}
    \right), \ 
    H_2 = \left(
    \begin{array}{cc}
        h_3^z & h_4 \\
        h_4 & h_3^x
    \end{array}
    \right),
\end{equation}
where 
\begin{align}
    h_{1}^{\alpha} & = 2 t_1^{\alpha} \left(\cos{}k_x+\cos{}k_y\right) + 2 t_3^{\alpha} \left[\cos(2k_x)+\cos(2k_y)\right] \nonumber \\ & + 2 t_2^{\alpha} \left[\cos(k_x+k_y)+\cos(k_x-k_y)\right] + \epsilon_{\alpha} \ (\alpha=z,x), \nonumber \\
    h_2 & = 2 t_3^{xz} \left(\cos{}k_x-\cos{}k_y\right) + 2 t_5^{xz} \left[\cos(2k_x)-\cos(2k_y)\right], \nonumber \\
    h_3^{\alpha} & = t_{\perp}^{\alpha} + 2 t_{\perp1}^{\alpha} \left(\cos{}k_x+\cos{}k_y\right) \nonumber \\ & + 2 t_{\perp2}^{\alpha} \left[\cos(k_x+k_y)+\cos(k_x-k_y)\right] \ (\alpha=z,x), \nonumber \\
    h_4 & = 2t_4^{xz} (\cos{}k_x-\cos{}k_y),
\end{align}
where $k_x$ and $k_y$ are the $x$ and $y$ components of $\bm{k}$, and the TB parameters are given in Table \ref{table_TB}. 

\begin{table}[htbp]
    \centering
    \caption{The parameters in the single-bilayer TB model. $t_n^{\alpha}$ represents the intralayer $n$-th nearest neighbor hopping of the $\alpha$ orbital electron. $t_{\perp}^{\alpha}$ represents the interlayer hopping of the $\alpha$ orbital electron. $t_{\perp{}n}^{\alpha}$ represents the interlayer $n$-th nearest neighbor hopping of the $\alpha$ orbital electron. $t_3^{xz}$, $t_4^{xz}$ and $t_5^{xz}$ represent the intralayer nearest neighbor hybridization, interlayer nearest neighbor hybridization and intralayer third nearest neighbor hybridization, respectively. $\epsilon_{\alpha}$ is the onsite energy of the $\alpha$ orbital. The unit of all parameters is eV. } \label{table_TB}
    \begin{tabular}{c|c|c|c|c|c}
        \hline
        \hline
        $t_1^z$&$t_2^z$&$t_3^z$&$t_{\perp}^z$&$t_{\perp1}^z$&$t_{\perp2}^z$\\
        \hline
        $-0.0889$&$-0.0167$&$-0.0147$&$-0.6408$&$0.0100$&$0.0104$\\
        \hline
        \hline
        $t_1^x$&$t_2^x$&$t_3^x$&$t_{\perp}^x$&$t_{\perp1}^x$&$t_{\perp2}^x$\\
        \hline
        $-0.4765$&$0.0772$&$-0.0613$&$0.0065$&$0$&$0$\\
        \hline
        \hline
        $t_3^{xz}$&$t_4^{xz}$&$t_5^{xz}$&$\epsilon_z$&$\epsilon_x$&\\
        \hline
        $0.2158$&$-0.0268$&$0.0273$&$9.0974$&$9.8963$&\\
        \hline
        \hline
    \end{tabular}
\end{table}

\subsection{The Standard Random Phase Approximation Approach}

In the RPA framework, the SC is driven by the spin or charge fluctuation whose propagator is given by the spin or charge susceptibility ($\chi^{(s/c)}$) renormalized up to the RPA level. 
When the interaction strength $U$ is larger than the critical interaction strength $U_c$, the spin or charge susceptibility will diverge and spin or charge order will form. In this system with repulsive interaction, the spin susceptibility diverges first as $U$ increases. 
When $U$ is below the critical interaction strength $U_c$, the critical temperature $T_c$ of the SC mediated by the spin fluctuation is determined by the pairing eigenvalue $\lambda$ through $T_c\sim e^{-1/\lambda}$, and the pairing symmetry is determined by the corresponding pairing eigenvector. 

\subsection{The Scanning Tunneling Microscopy Spectrum}

To calculate the STM spectrum, we start with the Matsubara Green's function
\begin{equation}
    \mathcal{G}(i,\tau) = - \sum_{\mu\alpha\sigma} \left\langle T_{\tau} c_{i\mu\alpha\sigma}(\tau) c^{\dagger}_{i\mu\alpha\sigma}(0) \right\rangle, 
\end{equation}
where $i$ is the site index, $\tau$ is the imaginary time, and $T_{\tau}$ represents the time-ordered product of the operators. Then we obtain the Matsubara Green's function in the frequency space by a Fourier transformation
\begin{equation}
    \mathcal{G}(i,\mathrm{i}\omega_n) = \int_{0}^{\beta} \mathrm{d}\tau \mathcal{G}(i,\tau) \mathrm{e}^{\mathrm{i}\omega_n\tau}, 
\end{equation}
where $\beta$ is the inverse temperature and $\omega_n=\frac{(2n+1)\pi}{\beta}$ is the frequency. Then we do an analytic continuation
\begin{equation}
    G^r(i,\omega+\mathrm{i}0^+) = \mathcal{G}(i,\mathrm{i}\omega_n\rightarrow\omega+\mathrm{i}0^+). 
\end{equation}
For the SC state at zero temperature, 
\begin{equation}
    G^r(i,\omega+\mathrm{i}0^+) = \sum_{\bm{k}m}\frac{u_{\bm{k}m}^2}{\omega-E_{\bm{k}m}+\mathrm{i}0^+}+\frac{v_{\bm{k}m}^2}{\omega+E_{\bm{k}m}+\mathrm{i}0^+}, 
\end{equation}
where $m$ is the band index, $E_{\bm{k}m}=\sqrt{\left(\varepsilon_{\bm{k}m}-\mu\right)^2+\Delta_{\bm{k}m}^2}$, $\varepsilon_{\bm{k}\alpha}$ is the $m$-th eigenvalue of the $\bm{k}$-space TB Hamiltonian $H_{\bm{k}}$, $\mu$ is the chemical potential, $\Delta_{\bm{k}m}$ is the gap function of the momentum $\bm{k}$ and band $m$, $u_{\bm{k}m}^2=\frac{1}{2}\left(1+\frac{\varepsilon_{\bm{k}m}-\mu}{E_{\bm{k}m}}\right)$ and $v_{\bm{k}m}^2=\frac{1}{2}\left(1-\frac{\varepsilon_{\bm{k}m}-\mu}{E_{\bm{k}m}}\right)$. Finally, the spectrum $A(\omega)$ is obtained from the imaginary part of $G^r$, 
\begin{equation}
    A(\omega) = - \frac{1}{\pi} \mathrm{Im} \left[ \sum_{\bm{k}m}\frac{u_{\bm{k}m}^2}{\omega-E_{\bm{k}m}+\mathrm{i}0^+}+\frac{v_{\bm{k}m}^2}{\omega+E_{\bm{k}m}+\mathrm{i}0^+} \right]. 
\end{equation}

\subsection{The Slave-boson-mean-field Approach}

The spin superexchange can be decoupled in the hopping and pairing channel, i.e., for the inter-layer one,
\begin{equation}
	\begin{aligned}
		&J_{\perp} \bm{S}_{t,\alpha}(i) \cdot \bm{S}_{b,\alpha}(i)   \\
		=&-
		\Big[\chi^{\alpha}_{\perp}
		\big(f_{t\alpha\uparrow}^{\dagger}(i) f_{b\alpha\uparrow}(i)
		+f_{t\alpha\downarrow}^{\dagger}(i) f_{b\alpha\downarrow}(i)\big)
		+\text{h.c.}- \frac{8|\chi^{\alpha}_{\perp}|^2}{3J_{\perp}}  \Big]\\ 
		&-
		\Big[\Delta^{\alpha}_{\perp}
		\big(f_{t\alpha\uparrow}^{\dagger}(i)f_{b\alpha\downarrow}^{\dagger} (i)
		-f_{t\alpha\downarrow}^{\dagger}(i)f_{b\alpha\uparrow}^{\dagger}(i) \big) 
		+\text{h.c.} -\frac{8|\Delta^{\alpha}_{\perp}|^2}{3J_{\perp}}   \Big],
	\end{aligned}
\end{equation}
At the mean-field level, order parameters are assumed to be site/layer independent.
The spinon hopping parameters are expressed as
\begin{equation}
\begin{aligned}\label{self_consistent}
\chi_{\parallel}^{\alpha}
&=\frac{3}{8} J_{\parallel} \langle f_{\mu\alpha\uparrow}^{\dagger}(i) f_{\mu\alpha\uparrow}(j)
+f_{\mu\alpha\downarrow}^{\dagger}(i) f_{\mu\alpha\downarrow}(j)\rangle ,   \\
\chi^{\alpha}_{\perp}
&=\frac{3}{8} J_{\perp} \langle f_{t\alpha\uparrow}^{\dagger}(i) f_{b\alpha\uparrow}(i)
+f_{t\alpha\downarrow}^{\dagger}(i) f_{b\alpha\downarrow}(i)\rangle, 
\end{aligned}
\end{equation}
respectively, where $\langle \cdots \rangle$ means the thermal average. The spinon pairing order parameters are expressed as
\begin{equation}
\begin{aligned}\label{self_consistent2}
\Delta_{\parallel}^{\alpha}
&=\frac{3}{8} J_{\parallel}\langle f_{\mu\alpha\downarrow}(j) f_{\mu\alpha\uparrow} (i)
-f_{\mu\alpha\uparrow}(j) f_{\mu\alpha\downarrow}(i) \rangle,  \\
\Delta^{\alpha}_{\perp}
&= \frac{3}{8} J_{\perp} \langle f_{t\alpha\downarrow}(i) f_{b\alpha\uparrow}(i) 
-f_{t\alpha\uparrow}(i) f_{b\alpha\downarrow}(i) \rangle,  
\end{aligned}
\end{equation}
respectively.

\bibliography{references}
\end{document}